\documentclass[11pt,twoside]{article}
\usepackage{asp2004}
\usepackage{psfig}
\usepackage{epsf}
\usepackage{graphics}
\usepackage{lscape}
\def\edcomment#1{\iffalse\marginpar{\raggedright\sl#1\/}\else\relax\fi}
\reversemarginpar

\begin{document}
\title{Far Ultraviolet Observations of the High Mass X-ray Binary 4U1700-37/HD153919}
\author{R.C. Iping, G. Sonneborn}
\affil{NASA's GSFC, code 681 Greenbelt MD 20771}
\author{L. Kaper, G. Hammerschlag-Hensberge}
\affil{University of Amsterdam, Kruislaan 403, 1098 SJ Amsterdam, the Netherlands}

\begin{abstract}
The high-mass  X-ray binary 4U1700-37/HD153919 has
 been observed with FUSE to study the effect of the X-ray source on 
the stellar wind of the primary. 
Phase dependent changes in the wind lines have been 
observed, indicating the creation of a X-ray ionization zone in the 
stellar wind. 
The X-ray luminosity of 4U1700-37 ionizes much of 
the wind and leaves a Stromgren zone. This disrupts the resonance-line acceleration 
of the wind in portions of the orbit, quenching the wind
 and disrupting the mass flow. This  effect 
was found for the first time in 4U1700-37.  This so-called Hatchett-
McCray (HM) effect had been predicted for 4U1700-37, but was not 
previously detected.

\end{abstract}
\thispagestyle{plain}

\section{Introduction 4U1700-37/HD153919}
High-mass X-ray binaries (HMXRB's) represent one of the key evolutionary paths for
massive stars and are unique laboratories of stellar evolution.
HMXRB's are critical to our understanding of supernovae
and the formation of neutron stars and black holes.  HD\,153919/4U1700-37 stands 
out as having the most massive and luminous primary of any HMXRB.

 4U1700-37 has an O6.5 Iaf primary in a 3.412-day orbital period with a neutron star
or black hole secondary (Rubin et al., 1996). The  primary eclipses the compact object for about 25\% of the binary period. 
The system is believed to have escaped from Sco OB I about 2.5 million years ago. 
The most recent mass estimates are those by Clark et al. (2002), who found 
$M_1=58 M_{\odot}$ and $M_2=2.4 M_{\odot}$.
No X-ray or radio pulses or other periodicities are known, however RXTE All Sky Monitor 
and CGRO BATSE data show evidence of a  13.8-day periodicity (Hong \& Hailey 2004).
The primary
slightly underfills its Roche lobe (Conti 1978), so 
accretion of the stellar wind onto the compact object powers the X-ray source.  
The X-ray light curve shows long eclipses and strong flaring behavior of which the physical origin
is not known.    X-ray pulsations have, however, not been detected, so that the
nature of the compact object is unclear.   As a consequence, the orbital and other fundamental
parameters of the system (e.g. masses of both components) usually derived from X-ray pulse timing analysis are not accurately known.

This HMXRB system is unique in showing Raman-scattered emission lines in its UV spectrum 
(Kaper et al. 1990).  These lines originate from strong EUV emission lines in the spectrum of the
X-ray source which are Raman-scattered to UV wavelengths by He II ions in the stellar wind.
In this way, emission lines in the X-ray spectrum can be detected in the UV.  
 Studies using IUE spectra of HD\,153919 have found two surprising properties of HD\,153919. 
There are a series of broad emission lines ($\Delta\lambda$ = 2-3 \AA) in the 200 \AA\ 
centered on about 1650 \AA\ that vary with orbital phase.  
These were shown to be Raman scattering of extreme UV photons ($\lambda \sim 255$ \AA) from 
the X-ray source by He II ions in the stellar wind (Kaper et al. 1990).  
The Raman lines appear and disappear with the binary phase, with maximum flux at 
phase 0.5 and minimum at phase 0.8.  Raman scattering,
a process in which the absorption of a photon is immediately followed by re-emission
of a photon at a different $\lambda$ (see Nussbaumer et al. 1989), 
is important in high temperature plasmas. [Rayleigh scattering is a special case of this process
 where the absorbed and emitted photons have the same $\lambda$.]
 
The Hatchett-McCray effect (Hatchett \& McCray 1977), the phase dependence of the 
stellar wind resonance lines due to X-ray ionization of the wind, has been observed in Vela X-1, 
SMC X-1, and other HMXRB.  The H-M effect was expected  for HD153919, but not observed 
(Kaper et al. 1993).  Similarly, no effect of the X-rays on the photospheric lines 
were found in IUE spectra (van Loon et al. 2001).
The stellar wind and O star UV spectrum appear to be unaffected by the X-ray 
source (Clark et al. 2002)

\section{FUSE Observations of 4U1700-37/HD153919}
The stellar wind diagnostics in the far ultraviolet (FUV) are a unique source of information
about the winds of massive stars.  Unlike the main wind lines in the UV longward of Lyman-$\alpha$ 
(N\,V 1238-42, Si\,IV 1394-1402 and C\,IV 1548-50), the wind features in the FUV probe higher
ionization states (O\,VI 1032-38) and the dominant ionization stages of the wind with species of 
lower cosmic abundance (S\,IV 1062-73, P\,V 1117-28) so the resonance lines are unsaturated 
(e.g. Massa et al. 2003).

4U1700-37 was observed by the Far Ultraviolet Spectroscopic Explorer (FUSE) at 
the four quadrature points of the 
binary orbit in 2003 April and August. The exposure times were $\sim4$ ksec at each orbital phase, 
producing high signal-to-noise spectra covering 900-1180 \AA\ with a spectral resolution
of 0.05\AA\ with S/N $>$30. 
The data were processed with latest FUSE calibration pipeline (CalFUSE v. 3.0.7). 
Figure 1 shows line profiles of 
several FUV wind lines on a velocity scale for two orbital phases, phase 0.49 
(X-ray source conjunction) and phase 0.71 (near elongation).  The S IV, P V, and Si IV lines are 
weakest at phase 0.49 and strongest at phase 0.71.  

Figure 2 further illustrates the phase dependence of the S IV 1062 line by plotting all four phases,
showing that 
these P-Cygni lines are strongest at the line center at phase 0.71 and weakest at phase 0.49. 
P\,V shows the identical behavior.  The C\,III 977 line is shown at the four orbital phases in Fig. 3.
C\,III is a trace species in the wind of an O 6.5 supergiant, since C\,IV is the dominant 
ionization stage of carbon.  Even so, much of the C\,III line profile is saturated (and also 
contaminated by interstellar Lyman-$\gamma$).  The high velocity edge of the C\,III profile is 
clearly visible, however, and is very similar to S\,IV.
This phase dependence is consistent with that predicted for the 
HM effect in 4U1700-37 by Hatchett \& McCray (1977).  It was never found in longer wavelength
lines because saturation masked the phase dependence.  The H-M effect 
has been observed for the first time in 4U1700-37, 27 years after its prediction.

\begin{figure}[!ht]
\plotfiddle{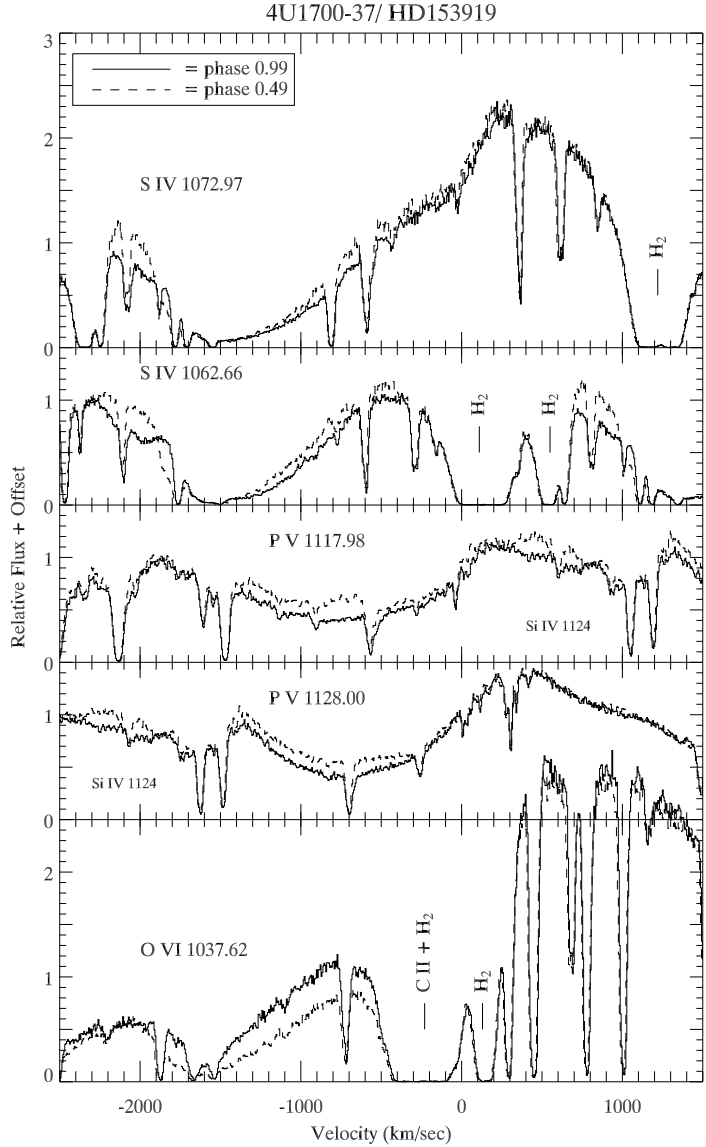}{5.9in}{0}{60}{60}{-200}{-20}
\caption{Stellar wind line profiles in 4U1700-37 at orbital phases 0.49 and 0.71 are shown
on a velocity scale.  
The narrow absorption features are interstellar, primarily Lyman band transitions
of H$_2$.}
\end{figure}

There is also evidence that the phase dependence of the different ions is a 
function of velocity in the wind.
The P\,V 1118-1128 and S\,IV 1063-1073 P- Cygni lines are weakest at $\phi=$ 0.49  
and strongest at $\phi= 0.71$ for the red wing and at 
$\phi= 0.0$ for the blue wing (see Figure 1). There is also a small variation in 
the terminal velocity.

\begin{figure}[!ht]
\plotone{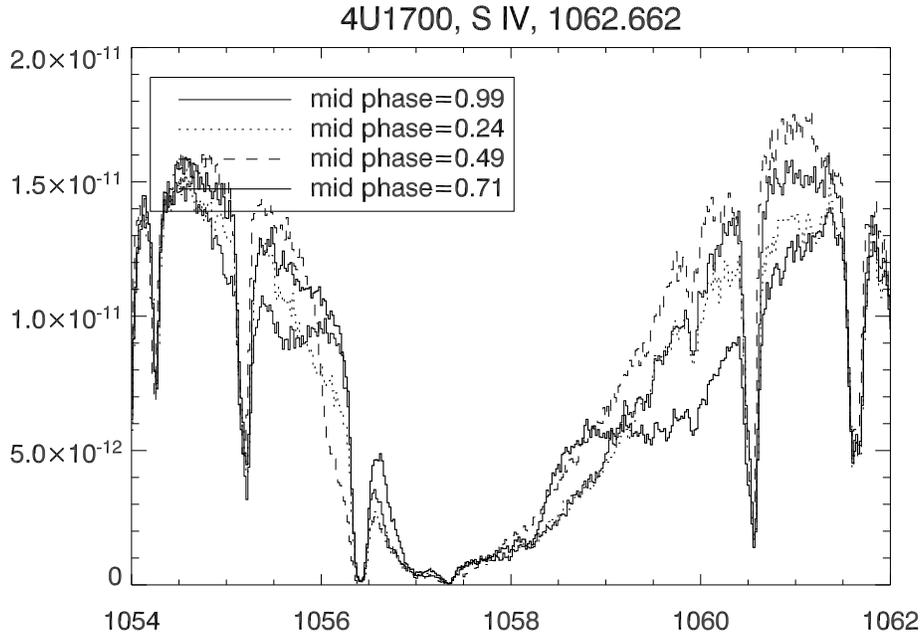}
\caption{P-Cygni profiles of the S\,IV 1062.662 line at the four quadrature 
points of the binary orbit.}
\end{figure}

\begin{figure}[!ht]
\plotone{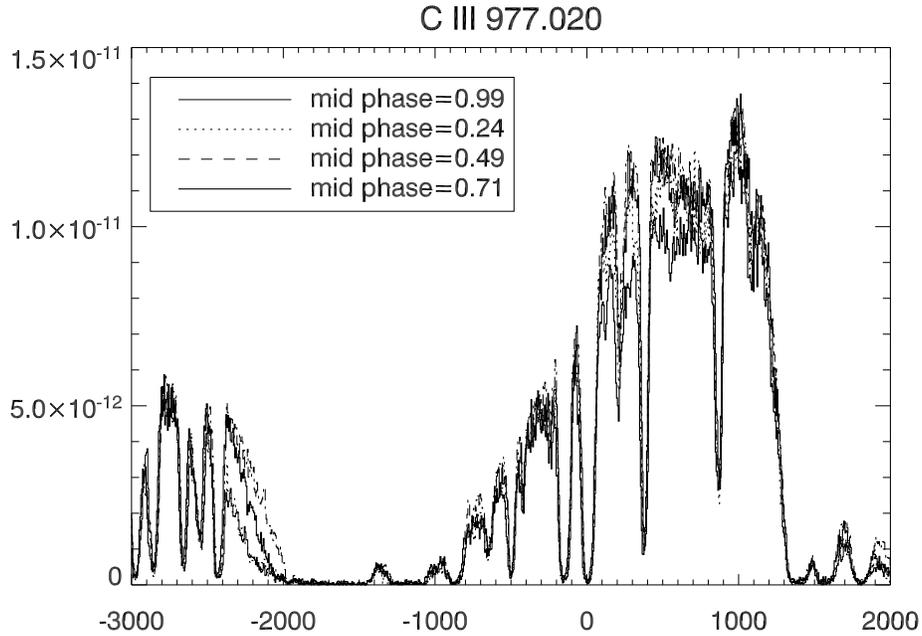}
\caption{The C III 977.020 resonance line is shown as a function of orbital phase.}
\end{figure}

The higher ionization lines, on the other hand, have a very different phase dependence. 
O\,VI 1038 and S\,VI 933-44 wind lines show orbital modulation different from  P\,V 
and S\,IV and are strongest at $\phi=0.5 $ and weakest at $\phi= 0.0$ 
(X-ray source eclipse), implying that O\,VI and S\,VI are byproducts of the wind's ionization by the
 X-ray source. Such variations were not observed in N\,V, Si\,IV and C\,IV because
 of their high optical depth.  The P\,V and S\,IV transitions, on the other hand, 
are excellent tracers of the ionization conditions in the O star's wind.  P\,V 
is the dominant ionization stage in the wind and has lower cosmic abundance than C, N or Si.
There is a small change in the terminal velocity ($v \sim 2000 \pm 200$ km/sec) 
of the wind  with phase.  The enhancement of O\,VI 
is not consistent with HM, and indicates
that the production of O$^{+5}$ ions is closely connected with proximity to the X-ray source.  
S\,IV and P\,V are removed from the wind at the same orbital phases as O\,VI is produced.
The modeling of the P-Cygni profiles with a modified version of the Sobolev
Exact Integration method, as described in van Loon et al. (2001), is in progress.

\end{document}